\documentstyle[12pt]{article}
\begin{document}
\baselineskip 22pt plus 2pt

\begin{center}
{\bf PANCHARATNAM AND BERRY PHASES IN} \\

{\bf THREE-LEVEL PHOTONIC SYSTEMS} \vspace{0.5cm}

Y. Ben-Aryeh \\ \ \\

Physics Department, Technion-Israel Institute of Technology \\
Haifa 32000, Israel \\ \ \\

e-mail:  phr65yb@physics.technion.ac.il; \\
Fax: \ 972-4-8221514
\end{center}

\noindent {\bf Abstract} \ \ A theoretical analysis of
Pancharatnam and Berry phases is made for biphoton three-level
systems, which are produced via frequency degenerate co-linear
spontaneous parametric down conversion (SPDC).  The general theory
of Pancharatnam phases is discussed with a special emphasis on
geodesic 'curves'in Hilbert space.  Explicit expressions for
Pancharatnam, dynamical and geometrical phases are derived for the
transformations produced by linear phase-converters.  The problem
of gauge invariance is treated along all the article.
\pagebreak

\noindent {\bf 1.  Introduction} \\ \ \\

An interesting development in quantum mechanics has been made by
Berry [1] who has shown that when a quantum-mechanical state is
developing around a closed circuit it obtains a geometrical phase
in addition to the well known dynamical phase.  Aharonov and
Anandan [2] have generalized the use of geometric phase to
non-adiabatic time developments.  Certain ideas from optics by
Pancharatnam [3] have been used for defining also phase changes
for partial and non unitary cycles [4,5].  The discovery of a
topological phase in quantum mechanics has led to a unified
approach to various topological phenomena in physics, both at the
quantum and classical levels.  The connections between the
geometric phase and fibre-bundle theories [6-10] have been
analyzed.  Following Berry's paper many experiments have been
developed for measuring the geometric part of the phase change, in
a variety of contexts, which are related to Lie group theories.
Most observations of Berry's phases are related to SU(2) and
Lorentz groups, and there is an enormous amount of literature on
such systems. In a previous paper [11] Berry's phases and
interference effects for three-level atoms, related to the SU(3)
algebra, have been analyzed, and the comparison between this work
and those of other authors have been discussed. In the present
article we would like to develop the theory of Pancharatnam and
Berry phases for three-level optical systems related to the SU(3)
algebra.  More specifically the theory of Pancharatnam phase is
developed here for a three-level photonic system which is realized
by the polarization states of the single mode biphoton field
[12-15].  A general review for the unitary evolution of polarized
light governed by the SU(2) group and its relation to topological
effects has been published [16].  We extend such treatment to
three-level photonic system. \\

There are close connections between the two concepts of
Pancharatnam and Berry phases. Since we are treating an optical
system we find it more useful to base our analysis on the concept
of Pancharatnam phase, but we would also like to show the
connection of our analysis with the concept of Berry phase. For
the basic understanding of our analysis for a specific system we
find it useful to review shortly the basic theories which are used
in our derivations. \\

The paper is arranged as follows:  In Section 2 we show the
relation between Berry phase and topology.  In Section 3 we
describe various properties of the Pancharatnam phase and its
relation to the Berry phase.  The main part of the paper is given
in Section 4, where we analyze possible interference effects which
can be obtained in single mode biphoton fields [12-15]. In Section
5 we summarize our results.
\vspace{0.5cm}

\noindent {\bf 2. Berry Phase Related to Topology} \\

When a quantum-mechanical (QM) state is evolving adiabatically in
time and transported around a circuit as an eigenstate of a
Hamiltonian with slowly varying parameters, it acquires a
geometrical phase factor in addition to the dynamical phase [1].
The geometrical phase factor is given by

\begin{eqnarray}
\gamma_n(T) = i \int^T_0 dt \langle n, \vec{R}(t)\mid d/dt \mid n,
\vec{R}(t)\rangle = i \oint d \vec{R}(t) \cdot \langle n,
\vec{R}(t)\mid \vec{\nabla}_{\vec{R}}
\mid n, \vec{R}(t)\rangle                  
\end{eqnarray}
for a Hermitian Hamiltonian $H(\vec{R}(t))$ with parameters
$R_i(t)$ which are slowly varying along a closed curve C, in
parameter space in time $T$. In the adiabatic approximation, it is
assumed that $\mid n, \vec{R}(t) \rangle$ is an eigenstate of the
Hamiltonian:
\begin{eqnarray}
H(\vec{R}(t)) \mid n, \vec{R} (t) \rangle = E_n \vec{R}(t)\mid n,
\vec{R} (t) \rangle                              
\end{eqnarray}
The phase factor acquired by the state $\mid \Psi_n(\vec{R},T)
\rangle$ consists of two parts:
\begin{eqnarray}
\mid \Psi_n(\vec{R},T)\rangle = \exp (i\gamma_n(T) \exp ( -i
\int^T_0 E_n(\tau)d\tau)\mid n, \vec{R} (t) \rangle   
\end{eqnarray}
where the first and second exponents on the right side of Eq. (3)
represent, respectively, the adiabatic Berry phase and the
dynamical phase. The term
\begin{eqnarray}
\vec{A}_{\vec{R}}(n,t) \equiv i \langle n,\vec{R}(t)\mid
\vec{\nabla}_{\vec{R}}\mid n,\vec{R} (t) \rangle   
\end{eqnarray}
is referred to as a vector potential.  As in electromagnetic
theory the vector potential is defined up to a gauge
transformation. By performing the transformation
\begin{eqnarray}
\mid n,\vec{R}(t) \rangle^\prime = \exp (i \alpha_n (\vec{R})
\rangle |n, \vec{R}(t) \rangle                                       
\end{eqnarray}
we induce a ``gauge transformation'' on $\vec{A}_{\vec{R}}(n,t)$:
\begin{eqnarray}
\vec{A}^\prime_{\vec{R}}(n,t)= \vec{A}_{\vec{R}}(n,t) -
\vec{\nabla}_{\vec{R}}(\alpha_n(\vec{R}))                   
\end{eqnarray}
>From the definition of Berry phase given by Eq. (1) it is clear
that the Berry phase is gauge invariant for a closed circuit. \\

Considering the state vector $\mid \Psi_n(\vec{R},t)\rangle$,
Aharonov and Anandan [2] removed its dynamical phase factor, by
defining a new state:
\begin{eqnarray}
|\phi_n(\vec{R},t) = \exp [i \int^t_0 h_n (\vec{R},t^\prime)
dt^\prime]|\Psi (n, \vec{R}(t)\rangle       
\end{eqnarray}
where
\begin{eqnarray}
h_n(\vec{R},t^\prime)= \langle \Psi_n (\vec{R},t^\prime)|\Psi_n
(\vec{R},t^\prime)\rangle^{-1} Re\langle \Psi_n (\vec{R},t^\prime)
|H(\vec{R},t^\prime)|\Psi_n (\vec{R},t^\prime)\rangle  
\end{eqnarray}
$|\phi_n (\vec{R},t^\prime)\rangle$ satisfies the equation:
\begin{eqnarray}
id/dt|\phi_n (\vec{R},t)\rangle = [H(t) - h_n (\vec{R},t)]|\phi_n
(\vec{R},t)\rangle                             
\end{eqnarray}
Multiplication of this equation on the left by $\langle \phi_n
(\vec{R},t)|$ yields the 'parallel transport' law:
\begin{eqnarray}
Im \langle \phi_n (\vec{R},t)|d/dt|\phi_n (\vec{R},t)\rangle = 0 
\end{eqnarray}

Following fibre-bundle theories [6-10] we refer to Eq. (10) as the
'connection'. The 'total space' $N$, in our analysis, includes all
normalized Schrodinger wavefunctions.  As is well known it is
possible to multiply the wavefunction by a phase factor $\exp
(i\theta)$, without changing the physical properties of the
quantum state.  All wavefunctions which differ only by a phase
factor are considered in QM as one ray.  The 'base space' $R$, in
the present analysis, includes all rays of Schrodinger
wavefunctions.  There is a projection $\pi : N \rightarrow R$ from
the total space to its base, i.e. from each wavefunction onto its
ray.  The 'fibre space' $F$ is related to our analysis by
including all wavefunctions which differ only by a phase factor as
one 'fibre'. \\

Let us assume, for simplicity, that an element of the 'bundle'
$|\phi_n (\vec{R},t)\rangle$ can be written
\begin{eqnarray}
|\phi_n (\vec{R},t)\rangle = \Omega |z_n (\vec{R},t)\rangle  
\end{eqnarray}
where $|z_n (\vec{R},t)\rangle$ is representative of a certain ray
and $\Omega$ is the phase factor, $\Omega = \exp (i\theta)$. On
the right side of Eq. (11) $\Omega$ represents the 'vertical'
part, which is along the 'fibre', while $|Z_n (\vec{R},t)\rangle$
represents the 'horizontal' part along the 'base'.  Substituting
Eq. (11) into the connection, Eq. (10), performing the derivative
and integrating this equation over a closed circuit during time
$T$ we have two contributions which are equal in magnitude and
opposite in sign, so that we obtain:
\begin{eqnarray}
{\rm Arg}(\Omega(T)) - \arg(\Omega(0)) \equiv \gamma_n(T) = i
\oint d\vec{R} \cdot \langle \phi_n
(\vec{R},t)|\vec{\nabla}_{\vec{R}}|\phi_n (\vec{R}(t)\rangle 
\end{eqnarray}
The geometrical phase factor is then given in Eq. (12) which is
the same relation as that of Eq. (1), but without any adiabatic
approximation.  One finds according to the above analysis that
although the quantum circuit is closed in the 'base space'
$(|z_n(T)\rangle = |z_n(0)\rangle )$ it is open in the 'total
space', as $|\phi_n(T) \rangle$ is different from $|\phi_n(0)
\rangle$, and this difference is given by the Berry phase.
\pagebreak

\noindent {\bf 3. Pancharatnam Phase and its Relation to
Topological Effects}
\vspace{0.25cm}

{\bf a. Definition of Pancharatnam Phase} \\

Consider two normalized nonorthogonal Hilbert states $|A \rangle$
and $|B \rangle$, and assume further that $|A \rangle$ is exposed
to $U(1)$ shift $e^{i\phi}$ [17].  The resulting interference
pattern is determined by
\begin{eqnarray}
I = |e^{i\phi}|A \rangle + |B \rangle |^2 = 2+2 |\langle A|B
\rangle | \cos [\phi - \arg \langle A|B \rangle ]   
\end{eqnarray}
where its maximum is obtained at the Pancharatnam relative phase
$\phi_0 \equiv \arg \langle A|B \rangle$.  This phase is reduced
to the $U(1)$ case whenever $|B \rangle = e^{i\theta}|A \rangle$
as it yields $\arg \langle A|B \rangle = \theta$.  In the original
treatments of the Berry phase [1,2]one considers a quantal system
evolving around a closed circuit, from an initial wavefunction $|A
\rangle$ to a final wavefunction $|B \rangle$ where $|B \rangle$
is obtained from $|A \rangle$ by a cyclic evolution, i.e., by
multiplication with a $U(1)$ phase factor.  Although the initial
phase of the quantal state (its 'fibre') is defined arbitrarily,
the phase difference between the state $|A \rangle$ and $|B
\rangle$ is well defined and can be observed by interferometric
methods. Pancharatnam provided the physical insight [17] that in a
noncyclic evolution if the final wave $|B \rangle$ is superimposed
on the initial state $|A \rangle$ only the component $\langle A|B
\rangle |A \rangle$ along $|A \rangle$ interferes with $|A
\rangle$. All other components of $|B \rangle$ which are
orthogonal to $|A \rangle$ merely add to the intensity, since
their cross terms with $|A \rangle$ vanish. The Pancharatnam
'connection' defines the phase between $|B \rangle$ and $|A
\rangle$ as $\phi_o = \arg \langle A|B \rangle$. The interference
amplitude $|\langle A|B\rangle |$ differs from unity if the
evolution is non-cyclic [17,18], and this difference leads to
reduction of the visibility in the interference pattern. The
Pancharatnam phase is indeterminate if $|B \rangle$ is orthogonal
to $|A \rangle$. Although Pancharatnam phase can be defined also
for mixed states [19-21] we restrict the analysis of the present
paper to pure states. The use of Berry and Pancharatnam phases to
noncyclic evolution raises the problem of gauge invariance which
is treated in the following paragraph. \vspace{0.25cm}

{\bf b. Pancharatnam Geometric and Dynamical Phases} \\

An extensive discussion of the mathematical and physical
properties of the Pancharatnam  phase has been presented in the
articles of Mukanda and Simon [22]. We describe here only a few
fundamental properties which are needed for understanding the
analysis given in Section 4 for a specific system. \\

Consider a one dimensional or one parameter 'curve' consisting of
a family of wave vectors $|\Psi(s)\rangle$ which is changing
continuously as a function of $s$.  $s$ is any parameter by which
$\Psi (s)$ is continuously changing (including time as a special
case). Assuming that $|\Psi(s)\rangle$ is a unit vector for all
$|\Psi(s)\rangle$ we get
\begin{eqnarray}
Re \langle \Psi(s)|\dot{\Psi}(s)\rangle = 0; \ \ \langle
\Psi(s)|\dot{\Psi}(s)\rangle = i \ Im \langle
\Psi(s)|\dot{\Psi}(s)\rangle                        
\end{eqnarray}
where the derivatives are according to the parameter $s$. One can
change the wavevectors $|\Psi(s)\rangle$ by a ``gauge
transformation'':
\begin{eqnarray}
|\Psi(s)\rangle \rightarrow |\Psi^\prime(s)\rangle =
e^{i\alpha(s)}|\Psi(s)\rangle; \ \ s\epsilon [s_1,s_2]   
\end{eqnarray}
where $\alpha (s)$ is a real smooth function of $s$ in the
interval $s\epsilon[s_1,s_2]$.  For cyclic transformation
described in the previous section the Berry phase is independent
of the gauge transformation. However, for noncyclic transformation
we get
\begin{eqnarray}
Im \langle \Psi^\prime(s)|\dot{\Psi}^\prime(s)\rangle = Im \langle
\Psi(s), \dot{\Psi}(s)\rangle + \dot{\alpha}(s)     
\end{eqnarray}
>From the above equations we find that we can construct a
functional in Hilbert space which is gauge invariant:
\begin{eqnarray}
&& \arg \langle \Psi^\prime(s_1), \Psi^\prime(s_2)\rangle - Im
\int^{s_2}_{s_1} ds \langle
\Psi^\prime(s)|\dot{\Psi}^\prime(s)\rangle = \\ \nonumber
&&\arg\langle \Psi(s_1), \Psi(s_2)\rangle - Im \
\int^{s_2}_{s_1}ds \langle \Psi(s)|\dot{\Psi}(s)\rangle = {\rm
gauge \ invariant}                   
\end{eqnarray}
This property can be interpreted according to the topological
description in the previous section. Here the wavefunction
$\Psi(s)\rangle$ includes its 'fibre' i.e. $|\Psi(s)\rangle$
belongs to the total space $N$ where it includes all wavefunctions
which differ in a phase factor $\exp (i\theta)$ [changing this
phase factor by the gauge transformation is considered in topology
as moving along the 'fibre'].  On the other hand, the functional
appearing in Eq. (17), which is gauge invariant represents a
projection from the total space to its basis and is defined as the
``geometric phase''. We find according to Eq. (17) that by
substracting the ``dynamical phase'', which is defined as
$Im\int^{s_2}_{s_1}ds \langle \Psi(s)|\dot{\Psi}(s)\rangle$, from
the Pancharatnam phase, which is defined here as $\arg\langle
\Psi(s_1), \Psi(s_2)\rangle$, we get the geometric phase which is
gauge invariant.  The interesting point here is that while Berry
phase analysis, as given in Section 2, is well defined (gauge
invariant) only for cyclic evolution, the geometric phase is well
defined also for noncyclic evolution (also for mixed states
[19-21]).  Eq. 17 can be reformulated as [22]:
\begin{eqnarray}
&& \phi_g(C) = \phi_p(N) - \phi_{\rm dyn}(N); \ \ \phi_p(N) = \arg
\{ \langle \Psi(s_1)|\Psi(s_2)\rangle \} ; \\ \nonumber &&
\phi_{\rm dyn}(N) = Im \int^{s_2}_{s_1}ds \langle
\Psi(s)|\dot{\Psi}(s)\rangle \ ,               
\end{eqnarray}
where $\phi_p(N)$ and $\phi_{\rm dyn}(N)$ are defined over the
total space $N$ and each of them is gauge dependent while
$\phi_g(C)$ is the geometric phase which is defined over the basis
space of rays and is gauge invariant.  Eqs. (17) and (18) have
been restricted to a dependence on only one parameter $s$, but
such restriction is valid for the optical system which will be
treated in Section 4. For a change of the wavefunction in an
interval $[s_1,s_2]$ one finds that the Pancharatnam phase is
fixed only by the initial and final wavefunctions
$|\Psi(s_1)\rangle$ and $|\Psi(s_2)\rangle$, respectively.  In
order to find the geometric phase one should find a way to
subtract the dynamical phase which is defined at each point $s$
and is obtained by its integration over the interval $[s_1,s_2]$.
\\

One can measure the Pancharatnam phase by applying Eq. (13), but
it might be quite difficult to subtract experimentally from it the
dynamical phase, although it can be done by using a theoretical
calculation. In considering consecutive wavefunctions 'curves' one
should take into account that the geometric and Pancharatram
phases do not have the additive property [19].  In comparison, the
dynamical phase of the total path can be obtained by adding the
dynamical phases of the 'curves' from which the total path is
composed.  Especially interesting are the geodesic 'curves' that
extremize the distance with respect to a certain metric. In the
next paragraph we introduce a metric for the present Hilbert space
and show how such extremum can be obtained. The geometric phase
which is obtained by subtracting the dynamical phase from the
Pancharatnam phase can have global properties while the geodesic
'curve' has also local differential properties. These properties
are obtained by extremizing a certain functional, leading to the
path with the shortest distance between two points in Hilbert
space. \vspace{0.25cm}

{\bf c. Hilbert Space Metric and Geodesics} \\

The distance between two quantum states $|\Psi(s)\rangle$ and
$|\Psi(s^\prime)\rangle$, representing a quantum system which is
developed according to parameter $s$, can be given as [23,24]
\begin{eqnarray}
dL^2 = 1 - \mid \langle \Psi(s)|\Psi(s^\prime)\rangle \mid^2 
\end{eqnarray}
As mentioned previously $s$ might represent not only the time $t$
but also any other parameter $s$ by which the wavefunction is
continuously developed.  The use of Eq. (19) is reasonable since
if the ray $|\Psi(s)\rangle$ is orthogonal to the ray
$|\Psi(s^\prime)\rangle$ then the distance between the two states
is equal to 1, while if $|\Psi(s^\prime)\rangle$ is equal to
$|\Psi(s)\rangle$ the distance vanishes.  Additional explanation
to this definition will be given also later, but let us see first
the mathematical derivations which are obtained from this
equation.\\ Assuming two close wavefunctions $|\Psi(s)\rangle$ and
$|\Psi(s+ds)\rangle$, then we get:
\begin{eqnarray}
\langle \Psi(s)|\Psi (s+ds)\rangle = 1 + ds \langle \Psi|
\frac{d}{ds} \Psi \rangle + \frac{1}{2} ds^2 \langle
\Psi|\frac{d^2}{ds^2}\Psi \rangle + 0 (ds^3)  
\end{eqnarray}
By using the second derivative of $\langle \Psi(s)|\Psi (s)
\langle$ we get
\begin{eqnarray}
\langle \Psi|\frac{d^2}{ds^2}\Psi \rangle + \langle
\frac{d^2}{ds^2}\Psi | \Psi \rangle + 2 \langle \frac{d}{ds}\Psi
| \frac{d}{ds} \Psi \rangle = 0 
\end{eqnarray}
By using Eqs. (19-21) one gets, after some algebra [23,24]:
\begin{eqnarray}
&& \left( \frac{dL}{ds} \right)^2 = \{ 1 - \langle \Psi(s)|\Psi
(s+ds)\rangle \langle \Psi (s+ds)|\Psi (s) \rangle \}/ds^2 \ \\
\nonumber && = \langle \frac{d\Psi}{ds} | \frac{d\Psi}{ds} \rangle
- \langle \frac{d\Psi}{ds}| \Psi \rangle \langle \Psi
|\frac{d\Psi}{ds} \rangle                             
\end{eqnarray}

The derivation of Eq. (22), which has been made by using the
definition (19), can be obtained also in a different way.  The
horizontal component of the tangent vector $d/ds | \Psi
(s)\rangle$ is given by $d/ds | \Psi (s)\rangle - \langle \Psi (s)
| \frac{d}{ds} \Psi (s) \rangle | \Psi (s) \rangle$. Here we have
subtracted from the derivative of the wavefunction its movement
along the 'fibre' [25], since it does not change the basis of the
wavefunction i.e., its ray. The norm of the above vector is given
after a straightforward algebra as:
\begin{eqnarray}
\left \{\langle \frac{d\Psi}{ds} |- \langle \frac{d}{ds}\Psi |\Psi
\rangle \langle \Psi | \right \} \left \{ \frac{d}{ds}| \Psi
\rangle -  \langle \Psi |\frac{d}{ds} \Psi \rangle | \Psi \rangle
\right \} = 2 (\frac{dL}{ds})^2  \nonumber
\end{eqnarray}
which is equivalent to Eq. (22) up to a multiplication by factor
2, which can be inserted arbitrarily in the definition (19). This
equation gives a physical insight for using Eq. (19). \\

In deriving Eq. (22) one neglects terms which are of order $ds^3$
or higher. Given the continuous 'curve' $C$ of the wavefunction
$\Psi (s)$ varying continuously as a function of the parameter $s$
from $s_1$ to $s_2$ one gets the functional [22,26]:
\begin{eqnarray}
L(C) = \int^{s_2}_{s_1} \left \{ \langle \dot{\Psi}(s)
|\dot{\Psi}(s) \rangle - \langle \Psi (s) |\dot{\Psi}(s) \rangle
\langle \dot{\Psi}(s), \Psi (s) \rangle \right \}^{1/2}   
\end{eqnarray}
The geodesic curve is obtained by extremizing this functional.
Mukanda and Simon [22] have shown that any 'curve' composed of a
normalized wavefunction which is changing continuously as a
function of $s$ and which obeys the equations:
\begin{eqnarray}
| \ddot{\Psi} (s) \rangle = - \langle \dot{\Psi}(s)|\dot{\Psi}(s)
\rangle | \Psi (s) \rangle \ ,                         
\end{eqnarray}
\begin{eqnarray}
\langle \Psi (s) |\dot{\Psi}(s) \rangle = 0 \ ,     
\end{eqnarray}
is a geodesic curve.  The path [22, 26-28]
\begin{eqnarray}
|\Psi(s)\rangle = |A \rangle \cos (s) + \left \{
\frac{|B\rangle-|A\rangle \langle A|B \rangle}{\sqrt{1-|\langle
B|A \rangle |^2}} \right \} \sin s                         
\end{eqnarray}
where $|A \rangle$ and $|B\rangle$ are the initial and final
wavefunctions is a geodesic curve as it fulfills Eqs. (24) and
(25). \\

The use of the geodesic equations (26) for noncyclic $SU(2)$
evolution has been discussed [27].  In a recent interesting
article [28] this equation has been applied for discussing
possible geometric phase measurement of three-level systems in
interferometry. According to Eq. (25) $| \Psi \rangle$ is
'parallel transported' [see Eq. (10)] as it leads to vanishing of
the dynamical phase at each point $s$.  The 'horizontal' property
given by Eq. (25) can be destroyed, {\bf but not the geodesic
property}, by multiplying the function $| \Psi (s) \rangle$ by
$e^{i\alpha (s)}$ where $\alpha (s)$ is any continuous function of
$s$. Such gauge transformation does not destroy the geodesic
property as a geodesic is gauge invariant  (any property of the
basis of the wavefunctions i.e., of the rays is gauge invariant).
\vspace{0.5cm}

{\bf d. The Vertex Theorem} \\

Let us assume that we have $N-1$ consecutive wavefunctions
'curves' where the r'th 'curve' is described by the transition
$\Psi_r \rightarrow \Psi_{r+1}$ and the total path is obtained by
summation over $r$ from 1 to $N-1$.  Using Eq. (18) we get:
\begin{eqnarray}
\phi_g(C) = \arg \langle \Psi_1| \Psi_N \rangle -
\sum^{N-1}_{r=1}\phi_{\rm dyn} (\Psi_r \rightarrow \Psi_{r+1})  
\end{eqnarray}
Although the Pancharatram and the dynamical phase can be gauge
dependent their difference given as $\phi_g(C)$ is gauge
independent. \\

For cases in which the geometric phase vanish in all the
transitions $\Psi_r \rightarrow \Psi_{r+1}$ we get:
\begin{eqnarray}
&& \phi_g(C) = \arg \langle \Psi_1| \Psi_N \rangle -
\sum^{N-1}_{r=1}\arg \langle \Psi_r|\Psi_{r+1}\rangle = \nonumber
\\[0.25cm]
&& - \arg \left\{ \langle \Psi_N| \Psi_1 \rangle \langle \Psi_1|
\Psi_2 \rangle \langle \Psi_2| \Psi_3 \rangle ... \langle
\Psi_{N-1}| \Psi_N \rangle \right \}
\end{eqnarray}
Eq. (28) can be considered as the general ``vertex theorem''. \\

Mukanda and Simon  [22] have proved, by developing certain
mathematical procedures, that any two normalized vectors in ray
space can be connected by a geodesic curve.  Therefore one can
apply Eq. (28) for the special case in which each of the
transitions $\Psi_r \rightarrow \Psi_{r+1}$ is a geodesic.  But in
deriving Eq. (28) we have assumed only that the geometric phase in
the ``global'' transition $\Psi_r \rightarrow \Psi_{r+1}$ vanish
and this does not necessarily imply that this transition is
geodesic. However, if we assume that the geometric phase vanish
continuously for each of the differential transition $\Psi (s)
\rightarrow \Psi (s + \varepsilon) \ (\varepsilon \rightarrow 0)$
in the interval $s\epsilon [s_1s_2]$ then the 'curve' is geodesic
and Eq. (28) gets the form [20]:
\begin{eqnarray}
&&\phi_g(C) = \nonumber \\
&& - \arg \{ \langle \Psi (s_2) |\Psi(s_1) \rangle \langle \Psi
(s_1) |\Psi(s_1+\varepsilon) \rangle \langle \Psi
(s_1+\varepsilon)| \Psi(s_1+2\varepsilon)\rangle ...  \nonumber \\
&& \langle \Psi
(s_1+(N-1)\varepsilon)| \Psi (s_2) \rangle \}                                                    
\end{eqnarray}
where $(N-1)\varepsilon = s_2-s_1\ , \ \ \ (N-1) \rightarrow
\infty \ , \ \ \varepsilon\rightarrow 0$.

Due to the fact that $\phi_g(C)$ is gauge invariant one can use
``parallel lift'' by which $\langle \Psi (s)|\dot{\Psi}(s) \rangle
= 0$ and then the vanishing of the geometric phase in each of the
intervals $(\Psi (s), \Psi (s+\varepsilon))$ implies $\langle \Psi
(s)|\Psi(s+\varepsilon) \rangle = 1$. Then we get the simple
expression
\begin{eqnarray}
\phi_g(C) = - \arg \langle \Psi (s_2) |\Psi(s_1) \rangle \ . 
\end{eqnarray}
For such cases one may measure $\phi_g(C)$ as a relative phase
shift in the interference pattern by applying Eq. (13), assuming
$|A \rangle = |\Psi (s_1) \rangle, |B \rangle = |\Psi (s_2)
\rangle$.  However, if the development is not given by `parallel
transport' one has to subtract the dynamical phase on the right
side of Eq. (30).
\pagebreak

\noindent {\bf 4.  Pancharatnam Phase for Polarized Biphotons} \\

{\bf a. Descriptions of the Biphoton States} \\

We consider a quantum system formed by two correlated photons - a
biphoton, emitted via frequency degenerate colinear spontaneous
parametric down conversion (SPDC) [12].  We assume that the
biphoton state can be described as
\begin{eqnarray}
|\Psi \rangle = c_1|2,0 \rangle + c_2|1,1 \rangle + c_3|0,2
\rangle                                
\end{eqnarray}
where $(N_x,N_y)$ means a state with $N_x$ photons in the
horizontal $(x)$ polarization mode and $N_y$ photons in the
vertical $(y)$ polarization mode, with $N_x+N_y=2$.  The states
$|2,0 \rangle$ and $|0,2 \rangle$ are generated via type-I SPDC
and the state $|1,1 \rangle$ via type-II SPDC.  Arbitrary
transformations of the polarization vectors $(c_1,c_2,c_3) \
[|c_1|^2+c_2|^2+|c_3|^2=1]$ are given by unitary $3\times 3$
matrix $G$, where $G^\dagger G=I$, det $G=1$, which form a
three-dimensional representation of the SU(3) group. \\

By passing from the basis $|2,0 \rangle,|1,1 \rangle,|0,2 \rangle$
to the basis
\begin{eqnarray}
|\Psi_+ \rangle = \frac{|2,0 \rangle + |0,2 \rangle}{\sqrt{2}}; \
\ |\Psi_- \rangle = \frac{|2,0 \rangle - |0,2 \rangle}{\sqrt{2}};
|\Psi_0 \rangle = |1,1 \rangle      
\end{eqnarray}
one obtains three states that can be transformed into one another
by means of only phase-plates.  It has been suggested to use
polarized biphotons as ternary analogs of two-state quantum
systems (qubits) [12,15].  Our aim in the present paper is,
however, different. We would like to study here possible
interference effects, by the use of biphotons which can be related
to Pancharatnam and Berry phases in a three-level photonic system.
The effect of the loss-free polarization converters (phase-plates)
on a biphoton state given by Eq. (31) has been described by
Burlakov and Klyshko [14].  For our purpose the effect of the
phase-plates is described by a $3 \times 3$ unitary matrix which
operates directly on the basis of states given by Eq. (32).  This
transformation matrix and its properties are described in the next
paragraph. \\

{\bf b. Transformations of Biphotons by Phase-Plates} \\

The effect of polarization converters on a biphoton given by Eq.
(31) as $(c_1,c_2,c_3)$ is described by the transformation matrix
[14]:
\begin{eqnarray}
G=\left(\begin{array}{ccc} t^2 & \sqrt{2}tr & r^2 \\[0.25cm]
-\sqrt{2}tr^* & |t|^2 - |r|^2 & \sqrt{2}t^* r \\[0.25cm] r^{*2} & -\sqrt{2}t^*r^*
& t^{*2} \end{array} \right)        
\end{eqnarray}
Here $t$ and $r$ are amplitude transmission and reflection
coefficients of a given converter.  We assume a linear phase-plate
with optical thickness $\delta$ and orientation $\chi$ relative to
the horizontal direction $x$ which corresponds to the
transformation $t = \cos \delta + i \sin \delta \cos (2\chi), \ r
= i \sin \delta \sin (2\chi)$. For a quarter-wave plate we have
$\delta = \pi/4$ and then $t = (1+i\cos (2\chi))/\sqrt{2}, \ r =
i\sin (2\chi))/\sqrt{2}$.  A half-wave plate gives $t = i \cos
(2\chi), \ r = i \sin (2\chi)$. \\

The basis of states $|2,0 \rangle, |11 \rangle$ and $|0,2 \rangle$
can be transformed to the basis of states $|\Psi_+ \rangle, \
|\Psi_- \rangle$ and $|\Psi_0 \rangle$ as
\begin{eqnarray}
\left(\begin{array}{c} \Psi_+ \\[0.25cm] \Psi_- \\[0.25cm] \Psi_0 \end{array} \right) =
A \left(\begin{array}{c} 2,0 \\[0.25cm]1,1
\\[0.25cm]0,2\end{array}\right)                     
\end{eqnarray}
where the matrix A is given by
\begin{eqnarray}
A = \left(\begin{array}{ccc}\frac{1}{\sqrt{2}} & 0 &
\frac{1}{\sqrt{2}}\\[0.25cm] \frac{1}{\sqrt{2}} & 0 & -\frac{1}{\sqrt{2}}
\\[0.25cm]0 & 1 & 0 \end{array}\right)      
\end{eqnarray}
The normalized biphoton state can be expressed in the new basis by
\begin{eqnarray}
| \Psi \rangle = d_1 | \Psi_+ \rangle + d_2 | \Psi_- \rangle + d_3
| \Psi_0 \rangle      
\end{eqnarray}
where the effect of the polarization converters on the biphoton of
Eq. (36) given as $(d_1,d_2,d_3)$ is described by the
transformation matrix: \\

\noindent $Q = AGA^{-1} =$
\begin{eqnarray}
\left(\begin{array}{ccc}\frac{t^2+r^2+r^{*2}+t^{*2}}{2} &
\frac{t^2-r^2+r^{*2}-t^{*2}}{2} & tr-t^*r^* \\[0.25cm]
\frac{t^2+r^2-r^{*2}-t^{*2}}{2} & \frac{t^2-r^2-r^{*2}+t^{*2}}{2}&
tr+t^*r^* \\[0.25cm] -tr^*+t^*r & -tr^*-t^*r & |t|^2-|r|^2
 \end{array}\right)                           
\end{eqnarray}
Assuming a linear phase-plate with optical thickness $\delta$ and
orientation $\chi$ the matrix $Q$ gets a more explicit form as:
\begin{eqnarray}
Q = \left(\begin{array}{ccc}\cos(2\delta) &
i\sin(2\delta)\cos(2\chi)
& i\sin\delta\sin(2\chi)\\[0.25cm]i\sin(2\delta)\cos(2\chi)&
\cos^2\delta-\sin^2\delta\cos (4\chi) & -\sin(4\chi)\sin^2\delta
\\[0.25cm]i\sin(2\delta)\sin(2\chi)& -\sin(4\chi)\sin^2(\delta) &
\cos^2\delta+\sin^2\delta \cos(4\chi) \end{array}\right)  
\end{eqnarray}
An interesting property of the matrix of Eqs. (37) and (38) is
that its diagonal matrix elements are real.  The relations between
the matrix $Q$ operating on the biphoton state $(d_1,d_2,d_3)$ and
Pancharatnam and Berry phases will be the subjects of the
following discussions. \\

{\bf c. Biphoton Dynamical and Pancharatnam Phases} \\

Following the general analysis presented in Section 3 the
Pancharatnam phase is developed as a function of the parameter $s$
and this is analogous to the development in time of Berry phase by
Schrodinger equation.  In order to apply the general theory to
phase converters we identify the parameter $s$ as the parameter
$\delta$(or later as 2$\delta$) in the transformation matrix $Q$
given by Eq. (38).  The input state vector is defined as
$(d_1,d_2,d_3)$ while the output state is obtained by multiplying
this input state by the matrix $Q$.  We assume that the initial
state is developed continuously by changing in the matrix $Q$ the
parameter $\delta$ from its initial value zero (giving a unit
matrix), to its final value $\delta_0$, which corresponds to the
optical depth of the converter.  By straightforward calculations
we find that the linear phase converter leads to a dynamical phase
shift given by:
\begin{eqnarray}
&& \int^{\delta_0}_0 d\delta (d^*_1,d^*_2,d^*_3) Q^{-1}\dot{Q}
\left(\begin{array}{c}d_1 \\[0.25cm]  d_2 \\[0.25cm]d_3 \\[0.25cm]
\end{array}\right) = \\[0.25cm]\nonumber
&& \int^{\delta_0}_0 d\delta \{ 2i\cos(2\chi)[d_1d^*_2+d^*_1d_2] +
2i\sin(2\chi)[d_1d^*_3+d^*_1d_3]\}        
\end{eqnarray}
The result obtained in Eq. (39) is quite simple showing that the
integrand in this integral is constant, depending only on the
initial state $(d_1,d_2,d_3)$ and the orientation $\chi$ which has
a fixed value for a certain converter.  The dynamical phase is
obtained by multiplying this integrand by $\delta_0$. \\

`Parallel transport' of the biphoton state vector is obtained by
the requirement that the integrand of Eq. (39) vanishes. `Parallel
transport' of the biphoton gives, however, trivial results since
it is easy to show that under this condition both the Pancharatnam
and geometric phases vanish.  This conclusion can be obtained by
calculating the imaginary value of $\langle \Psi_{\rm
in}|\Psi_{\rm out} \rangle$ where $|\Psi_{\rm in}\rangle$ is the
initial biphoton state vector while $|\Psi_{\rm out} \rangle$ is
obtained by multiplying this initial state by the transformation
matrix $Q$.  We get:
\begin{eqnarray}
&&Im \langle\Psi_{in}|\Psi_{out}\rangle = Im \{(d^*_1,d^*_2,d^*_3)
Q \left(\begin{array}{c}d_1 \\[0.25cm] d_2 \\[0.25cm] d_3\end{array}\right)
\} = \\[0.25cm]\nonumber
&& \sin 2\delta \{\cos 2\chi (d^*_1 d_2+d^*_2d_1) + \sin (2\chi)
(d^*_1 d_3+d_1d^*_3)\}               
\end{eqnarray}
Therefore, vanishing of the integrand in Eq. (39), which is the
condition for `parallel transport' implies vanishing also of the
Pancharatnam phase.  Cases for which the total path is composed by
a continuum of differential `curves' elements, where for each of
them the geometric phase vanishes while the geometric phase for
the total path might be different from zero can be defined as
geodesic curves in its most general meaning.  For an infinitesimal
evolution of the biphoton state from an initial state
$(d_1,d_2,d_3)$, obtained by using additional thickness $d\delta$,
the change in geometric phase vanishes.  This conclusion can be
obtained by using in Eq. (40) the approximation $\sin
(2\delta)\simeq 2\delta$ and compare the Pancharatnam phase change
with that obtained for the dynamical phase change given by Eq.
(39). Therefore, the general evolution of the Biphoton state by
phase converters can be considered as geodesic curves in a general
sense.  More basic and restricted definitions of geodesic 'curves'
will be given later in the article. \\

There are other interesting effects which can be observed by
linear phase converters: \\
1) \ By calculating the eigenvalues and eigenvectors of the
transformation matrix $Q$ one can describe cyclic transformations
by which the initial state vector is multiplied by $U(1)$ phase
factor. \\
2) \ Special geodesic developments can be obtained for biphoton
states under special conditions, which can lead to interesting
interference effects. \\
3) \ Under general conditions, the geometric phase can be
calculated by using the general expressions of Eq. (18). \\

{\bf d. Eigenvalues and Eigenvectors of the Transformation Matrix} \\

Since the transformation matrix $Q$ is unitary its eigenvalues are
given by complex numbers with a unit absolute value. By
calculating the eigenvectors of the matrix $Q$ one finds the
initial biphoton states for which the transformation $Q$ produces
cyclic transformations i.e., the initial state vectors are
multiplied by these complex numbers. The general calculation of
the eigenvalues of the matrix $Q$ for arbitrary optical depth
$\delta$ leads to cubic equations.  For simplicity, we demonstrate
the general procedure by applying it to relatively simple cases.
\\

For a quarter-wave plate with $2\delta = \pi/2$ the determinant
for the eigenvalues $\lambda$ is given by
\begin{eqnarray}
-\lambda^3 + \lambda^2 - \lambda + 1 = 0   
\end{eqnarray}
The roots of Eq. (41) are given by $\lambda = \pm i,1$
(independent of the orientation $\chi$). For the eigenvalue
$\lambda = i$ the normalized eigenvector is given by
\begin{eqnarray}
d_1 = \frac{1}{\sqrt{2}}; \ \ d_2 =  \frac{\cos
(2\chi)}{\sqrt{2}}; \
\ d_3 =  \frac{\sin (2\chi)}{\sqrt{2}}      
\end{eqnarray}
Here we have assumed that $d_1$ is real and this can be done since
the phase of the initial biphoton state is arbitrary. However, the
phase factor $e^{i\pi/2}$ by which the initial state is multiplied
for getting the final state vector is well defined and can be
observed in interference experiments.  The dynamical phase can be
calculated for the initial state, given by Eq. (42) with $2\delta
= \pi/2$, by Eq. (39) as:
\begin{eqnarray}
\int^{\pi/4}_0 d\delta \{\cos^2(2\chi) + 2i \sin^2 (2\chi) \} = i
\pi/2                                  
\end{eqnarray}
So we find that all the Pancharatnam phase is a dynamical phase.
Similar calculations can be given for the roots $\lambda = - i$
and 1. \\

For a half-wave plate the straightforward calculations show that
the roots are given by $ = \pm 1$.  This result is obvious since
the half-wave plate can be considered as the operation of two
consecutive quarter-wave plates with the same orientation $\chi$,
so that the roots for the half-wave plate are obtained by the
squares of the roots of the quarter-wave plates.  One finds that
the roots for phase-plates with arbitrary optical depth $\delta$
will be given by $e^{\pm i2\delta}, \ 1$. One can also calculate
the eigenvalues and the eigenvectors for multiplication of two (or
more) phase plates with different orientations $\chi$.  Our
interest, however, is to calculate geometric phases for noncyclic
developments and for this purpose geodesic developments are
described in the next paragraph, for special cases. \\

{\bf e. \ Geodesic Development of Biphoton States} \\

Let us assume a phase converter with orientation given by $\cos
(2\chi)=1$.  We assume also that this phase converter transform a
special initial biphoton state for which $d_3 = 0 \ (|d_1|^2 +
|d_2|^2 = 1$). \ We can easily prove that this biphoton state is
developed as a function of $\delta$ along a geodesic 'curve', as
the geodesic equations (24) is obeyed.  Defining $s = 2\delta$ we
obtain

\begin{eqnarray}
\langle \dot{\Psi}(s)|\dot{\Psi}(s)\rangle = (d^*_1,d^*_2) \left(
\begin{array}{cc}-\sin s & -i\cos s \\[0.25cm]-i\cos s &
-\sin s \end{array}\right)\left(
\begin{array}{cc}-\sin s & i\cos s \\[0.25cm]i\cos s &
-\sin s \end{array}\right)\left(
\begin{array}
{c}d_1   \\[0.25cm]d_2
 \end{array}\right) = 1  \nonumber \\
 ~~~~~~~~      
\end{eqnarray}
and as $|\ddot{\Psi}(s)\rangle = - | \Psi (s)\rangle$, equation
(24) is fulfilled. Although this biphoton does not obey the
'parallel transport' of Eq. (25) we can use a gauge transformation
of $|\Psi (s)\rangle$ which will lead to a 'parallel transport'.
This gauge transformation can be given according to Eq. (39) by
\begin{eqnarray}
|\Psi (s) \rangle \rightarrow |\Psi^\prime(s) \rangle = |\Psi (s)
\rangle \exp \{ -is [d_1d^*_2+d^*_1d_2] \}     
\end{eqnarray}
The conclusion from this transformation is that for the present
biphoton state $|\Psi (s) \rangle$ the geodesic property is
preserved although the 'horizontal' property is destroyed.  [see
the discussions after Eq. (26)]. Since the 'horizontal' property
is not preserved one can use Eq. (30) but the dynamical phase
should be subtracted from the right side of this equation. One
should therefore notice that the geodesic 'curve' described here
is basically different from that suggested in Ref. [28]. \\

For the present geodesic 'curve' we find that $\theta = \arg
\langle \Psi (s_1)|\Psi (s_2)\rangle$ is given as a function of
the optical depth $2\delta = s$ by
\begin{eqnarray}
\tan \theta = \tan (s) (d^*_1d_2 + d^*_2 d_1)    
\end{eqnarray}
and the geometric phase is given by
\begin{eqnarray}
\phi_g(C) = \theta - s (d^*_1d_2 + d^*_2 d_1) 
\end{eqnarray}
According to Eq. (46) $\theta = s = 2\delta$ only for the special
cases for which $d^*_1d_2 + d^*_2 d_1 = 1$, and only under this
condition the geometric phase vanishes.  For more general cases we
find that when $s$ is changing from $\pi/2 - \varepsilon \
(\varepsilon \rightarrow 0)$ to $\pi/2 + \varepsilon$ there is a
jump in the geometric
phase of $\pi$. Such phase jumps can be observed [16]. \\

In a similar way to the above analysis one can obtain another
geodesic 'curve' by using a phase converter with orientation given
by $\sin (2\chi) = 1$ and an initial state for which $d_2 = 0 \
(|d_1|^2 + |d_3|^2 = 1)$.  The present geodesic 'curves' are
obtained by the development of only two levels chosen from the
three-level system. \\

In the next paragraph we demonstrate calculations of the geometric
phase for a real three-level system. We also explain the
difference between the general biphotons transformation which we
defined as geodesic 'curves' in the more general sense and the
more restricted definition given by Eq. (24). \\

{\bf f. Geometric Phase Obtained by Linear Converters} \\

By the derivations of Eqs. (39) and (40) we implied that the
linear converters produce 'curves' which are geodesic in a general
sense.  Since the transformation of the biphoton state by the
matrix $Q$ of Eq. (38) does not obey the geodesic Eq. (24) we need
to justify our assumption also from the mathematical point of
view. The geodesic equation (24) is basically a harmonic
oscillator equation for the wavefunction $| \Psi (s) \rangle$.  We
find according to Eqs. (39) and (40) that only the imaginary
elements of the matrix $Q$ contribute to Pancharatnam and
dynamical phases.  The imaginary part of the $Q$ matrix fulfills
the equation
\begin{eqnarray}
Im (\ddot{Q} + 4Q) = 0       
\end{eqnarray}
which is basically an harmonic oscillator equation explaining the
geodesic property of this transformation in its general sense. \\

Let us demonstrate the use of Eqs. (18) for two consecutive
transformations operating on an initial state $(d_1,d_2,d_3)$,
first by using quarter-wave plate with an orientation $\chi=0$ and
second by using half-wave plate with orientation $\chi = \pi/4$.
We define the Pancharatnam phase in the first, second and total
transformations as $\theta^{(1)}_{Pan}, \theta^{(2)}_{Pan}$, and
$\theta^{(3)}_{Pan}$, respectively. Dynamical phases in the first,
second and total transformations are defined as
$\theta^{(1)}_{dyn}, \theta^{(2)}_{dyn}$ and $\theta^{(3)}_{dyn}$,
respectively. By straightforward calculations we get:
\begin{eqnarray}
&& \tan \theta^{(1)}_{Pan} =
\frac{d^*_1d_2+d^*_2d_1}{|d_1|^2+|d_2|^2 + |d_3|^2 \sqrt{2}} \ ; \
\theta^{(2)}_{Pan} = \theta^{(tot)}_{Pan} = 0 \ ; \nonumber \\
&& \theta^{(2)}_{dyn} = 2\pi \{(d^*_1d_3+d_1d^*_3) + i (d_2d^*_3 -
d^*_2d_3) \} \ ; \nonumber \\
&& \theta^{(1)}_{dyn}= \pi (d^*_1d_2+d^*_2d_1) \ ; \nonumber \\
&&
   \theta_{dyn}^{(tot)} = \theta^{(1)}_{dyn} + \theta_{dyn}^{(2)} 
\end{eqnarray}
We find that the total Pancharatnam phase vanishes and therefore
in the present case the total geometric phase is equal to minus
the total dynamical phase.
\vspace{0.5cm}

\noindent {\bf 5. Summary and Discussion} \\

The present paper has analyzed the transformations of biphoton
states which can be obtained by the use of linear phase converters
in relation to Pancharatnam and Berry phases. While most of the
previous works have treated two-level systems we analyze here a
special three-level photonic system. In order to understand the
present analysis the basic concept of Pancharatnam phase is
explained in relation to topological effects.  The fundamental
properties of Pancharatnam and Berry phases are reviewed for the
purpose of using them in the analysis of a specific system. \\

In Section 2 the relation between Berry phase and topology is
explained. The phase obtained by an atomic system developing
according to Schrodinger equation can be separated into a
dynamical and geometric phase. This topological separation is
common for Berry and Pancharatnam phase effects in both atomic and
photonic systems. The Berry phase is usually calculated for a
closed circuit for which the calculation of phase is gauge
invariant. An important development has been made by the use of
Pancharatnam phase which gives, after the subtraction of the
dynamical phase, a geometric phase which is gauge invariant. \\

In Section 3 the main properties of the Pancharatnam phase are
treated.  The basic equation for obtaining the geometric phase in
any atomic or photonic system is given by Eq. (18).  A special
emphasis is made in the present work on geodesic 'curves' which
are related to Hilbert space metric.  The basic equations for
geodesics and for 'parallel transport' are given in Eqs. (24) and
(25) respectively. A vertex theorem is developed for 'parallel
transport' of a Hilbert state along a geodesic 'curve'. \\

The main results of the present work are given in Section 4 for
analyzing Pancharatnam phases obtained by the transformations of
biphotons using phase-plates.  We have used the basis of states
$\Psi_+, \Psi_-$ and $\Psi_0$ given by Eq. (34) and the general
transformation of these states by the phase-plates is given by Eq.
(38). A general formula for the dynamical phase obtained by this
transformation is given by Eq. (39). In Eq. (40) we have obtained
the result for $Im \langle \Psi_{\rm in}|\Psi_{\rm out} \rangle$
and have shown that for an infinitesimal transformation the
geometric phase vanishes.  We find therefore that the
transformation given by the matrix $Q$ of Eq. (38) produces a
geodesic curve in its general meaning. Cyclic transformations for
the Pancharatnam phases have been obtained by the use of
eigenvalues and eigenvectors of the transformation matrix but such
transformations give only dynamical phases. Geodesic 'curves'
fulfilling the geodesic equation (24) are obtained for special
transformations.  For such geodesic 'curves' the 'horizontal'
property of the 'parallel transport' is destroyed but not the
geodesic property. Possible phase jumps of $\pi$ are related to a
certain discontinuity in the geometric phase. Although the
geodesic equation (24) is fulfilled only for special cases,
describing the development of two levels out of the three-level
system, it has been shown that the general transformation of the
three optical levels by the matrix $Q$ is also geodesic in its
general sense. Such geodesic 'curves' are obtained for vanishing
geometrical phases for small changes in the transformation matrix,
but obtained as a global geometric phase change in the total
curve, related to non-additivity of the geometrical phase. This
geodesic property has been justified mathematically also by
replacing the geodesic equation (24) by the present geodesic
equation (48).  In this new equation only the imaginary part of
the $Q$ matrix is taken into account, since the real part of the
$Q$ matrix does not contribute to Pancharatnam and geometric
phases.  The use of the general equation (18) for calculating
geometrical phases is demonstrated by a calculation for a special
case. \vspace{0.5cm}

\noindent{\bf Acknowledgement} \\

The author would like to thank S.P. Kulik for interesting
discussions.
\pagebreak

\noindent {\bf References}

\begin{enumerate}
\item Berry, M.V., 1984, Proc. R. Soc. A. {\bf 392}, 45-57.
\item Aharonov, Y. and Anandan, J., 1987, Phys. Rev. Lett. {\bf 58},
1593-1596.
\item Pancharatnam, S., 1956, Proc. Ind. Acad. Sci. A {\bf 44},
247-262.
\item Jordan, T.F., 1988, Phys. Rev. A. {\bf 38}, 1590-1592.
\item Samuel, J. and Bhandari, R., 1988, Phys. Rev. Lett. {\bf
60}, 2339-2342.
\item Eguchi, T., Gilkey, P.B. and Hanson, A.G., 1980, Physics
Rep. {\bf 66}, 213-393.
\item Chern, S.S., Chen, W.H. and Lam, K.S., 1998, {\it Lectures on
Differential Geometry} (Singapore: World Scientific).
\item Kobayashi, S. and Nomizu, K., 1969, {\it Foundations of
Differential Geometry} (New York: Interscience).
\item Nash, C. and Sen, S., 1983, {\it Topology and Geometry for
Physicists} (London: Academic Press).
\item Simon, B., 1983, Phys. Rev. Lett. {\bf 51}, 2167-2170.
\item Ben Aryeh, Y., 2002, J. Mod. Optics, {\bf 49}, 207-220.
\item Burlakov, A.V., Chekova, M.V., Karabutova, O.A., Klyshko,
D.N. and Kulik, S.P., 1999, Phys. Rev. A, {\bf 60}, R4209-R4212.
\item Burlakov, A.V., Chekova, M.V., Karabutova, O.A. and Kulik,
S.P., 2001, Phys. Rev. A, {\bf 64}, 041803, 1-4.
\item Burlakov, A.V. and Klyshko, D.N., 1999, JETP Lett. {\bf 69},
839-843.
\item Burlakov, A.V., Krivitskiy, L.A., Kulik, S.P., Maslennikov,
G.A. and Chekova M.V., July 2002, arxiv:quant-ph/0207096.
\item Bhandari, R., 1997, Phys. Rep., {\bf 281}, 1-64.
\item Wagh, A.G. and Rakhecha, V.C., 1995, Phys. Lett. A., {\bf
197}, 107-111.
\item Sjoqvist, E., 2001, Phys. Lett. A. {\bf 286}, 4-6.
\item Sjoqvist, E., Feb. 2002, arxiv: quant-ph/0202078.
\item Sjoqvist, E., Pati, A.K., Ekert, A., Anandan, J.S.,
Ericsson, M., Oi, D.K.L. and Vederal, V., 2000, Phys. Rev. Lett.,
2845-2849.
\item Ericsson, M., Pati, A.K., Sjoqvist, E., Brannlund, J. and
Oi, D.K.L., June 2002, arxiv: quant-ph/0206063.
\item Mukanda, N. and Simon, R., 1993, Annals of Physics, {\bf
228}, 205-340.
\item Anandan, J. and Aharonov, Y., 1990, Phys. Rev. Lett. {\bf
65}, 1697-1700.
\item Anandan, J., 1991, Foundations of Physics {\bf 21},
1265-1284.
\item Bohm, A., Boya, L.J. and Kendrik, B., 1991, Phys. Rev. A
{\bf 43}, 1206-1210.
\item Arvind, Malesh, K.S. and Mukanda, N., 1997, J. Phys. A:
Math. Gen. {\bf 30}, 2417-2431.
\item Sjoqvist, E., 2001, Phys. Rev. A, {\bf 63}, 035602, 1-4.
\item Sanders, B.C., De Guise, H., Bartlett, S.D. and Zhang, W.,
2001, Phys. Rev. Lett. {\bf 86}, 369-372.
\end{enumerate}

\end{document}